# Correlated topological-polarization surface states in the narrow-gap insulator $FeSb_2$

Takahiro Iwagaki[1,2,†], Hideki Matsuoka[1], Ginta Hoshino[3], Kanata Watanabe[3], Shungo Aoyagi[1,2], Shunsuke Kitou[4], Yuiga Nakamura[5], Motoaki Hirayama[6], Takashi Koretsune[3] and Naoya Kanazawa[1 †]

[1] *Institute of Industrial Science, The University of Tokyo, Tokyo 153-8505, Japan*
[2] *Department of Applied Physics, The University of Tokyo, Tokyo 113-8656, Japan*
[3] *Department of Physics, Tohoku University, Sendai, Miyagi 980-8578, Japan*
[4] *Department of Advanced Materials Science, The University of Tokyo, Kashiwa, Chiba 277-8561, Japan*
[5] *Japan Synchrotron Radiation Research Institute (JASRI), SPring-8, Hyogo 679-5198, Japan*
[6] *RIKEN Center for Emergent Matter Science, Wako 351-0198, Japan*

† To whom correspondence should be addressed. E-mail: iwagaki@iis.u-tokyo.ac.jp, naoya-k@iis.u-tokyo.ac.jp

**Abstract**

Strong electron correlations and band topology each generate rich quantum phases, but conflicting elemental requirements have largely kept them apart. Topological polarization offers a route to unite them, producing polar surface states from bonding charge without spin-orbit coupling and thereby extending band topology to correlated 3*d* transition-metal compounds. Here we demonstrate that epitaxial thin films of the narrow-gap insulator $FeSb_2$ host metallic polar surface states of topological-polarization origin, governed by the strong correlations of the bulk. Nonreciprocal surface transport emerges only below the onset temperature of a correlation-driven reconstruction of the bulk Fe 3*d* orbital occupation, providing direct evidence of bulk-edge correspondence in a correlated topological system. Moreover, electrostatic gating drives this correlated surface across a quantum phase transition into a ferromagnetic or possibly altermagnetic state. Our results establish topological polarization as a design principle for correlated topological phases in a broad range of materials.

## Main text

The interplay between strong electron correlations and band topology[1–7] has become a frontier of condensed matter physics. When the two cooperate, they generate striking quantum phases, including fractional Chern insulators (*e.g.*, twisted bilayer $MoTe_2$)[8–12], topological Kondo insulators (*e.g.*, $SmB_6$)[4,13] and Weyl-Kondo semimetals (*e.g.*, $Ce_3Bi_4Pd_3$)[5,14]. This intersection holds the promise of further unexplored quantum states and functionalities.

Combining correlation and topology in a single material is fundamentally difficult. Berry-curvature-based topology generally relies on band inversion driven by strong spin-orbit coupling (SOC), which favors heavy elements with extended 6*p* orbitals. In contrast, strong correlations are most pronounced in the localized orbitals of 3*d* transition metals with large Coulomb repulsion. These conflicting elemental requirements have kept the coexistence of correlation and topology limited to a few systems.

Recent developments in each field have relaxed this elemental constraint from both sides. On the correlation side, strong correlations have long been confined mostly to 3*d* oxides, where ionic bonding isolates the narrow 3*d* bands. In intermetallic compounds, by contrast, covalent bonding broadens the bands and quenches correlations. Flat bands lift this constraint[2,15]. They can be engineered by controlling the electronic hopping, through the twist angle in moiré superlattices[16], destructive interference in kagomé lattices[17], or anisotropic hopping set by the crystal geometry. The suppressed bandwidth allows the on-site Coulomb repulsion to dominate, driving a wide range of systems, from moiré superlattices to intermetallic compounds, into the correlated regime. On the topology side, topological surface states have long demanded the heavy elements required for band inversion. Topological polarization lifts this constraint. It is captured by the Zak phase[18–21] and classified by the theory of obstructed atomic insulators[22–25], yielding surface states without SOC-driven band inversion. The Zak phase, a geometric phase of the bulk bands, corresponds to the position of the valence Wannier centers. When it is nonzero (quantized to π by inversion symmetry), these centers are displaced from the atoms into the interatomic region. Terminating the crystal then exposes them as polar surface states[26–28]. Because this polarization arises from the bonding charge rather than from SOC, it extends topological surface states to light-element compounds. Unlike the Dirac states of topological insulators, the polar surface states are not tied to a linear dispersion.

These two advances have developed in parallel but have rarely been brought together. If such a topological-polarization surface state emerges in a strongly correlated material, the surface should inherit the hallmark of strong correlations: competing quantum phases tunable by external fields. It remains an open challenge to realize such a correlated topological surface and to drive it across a quantum phase transition.

Here we realize a correlated topological surface in the 3*d* narrow-gap insulator $FeSb_2$. Combining multiple theoretical and experimental approaches, we demonstrate that strong correlations govern a topological-polarization surface state and can drive it across a quantum phase transition. First-principles calculations identify a topological polarization and the resulting polar surface states. Synchrotron X-ray diffraction, together with density functional theory plus dynamical mean-field theory (DFT+DMFT) calculations[29–31], shows that strong correlations reconstruct the orbital occupation with increasing temperature, observed as a marked change in the valence electron density (VED). Surface transport exhibits a nonreciprocal response, characteristic of a polar surface and forbidden in the centrosymmetric bulk. This response emerges only below the temperature scale of the bulk orbital reconstruction, which weakens the topological polarization. This concurrence demonstrates that the surface state directly reflects the bulk electronic state through a bulk-edge correspondence. In electric-double-layer transistors (EDLTs), a gate voltage induces a hysteretic anomalous Hall effect (AHE) at the lowest temperatures, indicating a quantum phase transition into a surface ferromagnetic (or possibly altermagnetic) state.

**First-principles calculations on bulk and surface electronic states in $FeSb_2$**

$FeSb_2$ crystallizes in the centrosymmetric marcasite structure with space group *Pnnm* (Fig. 1**a**). It is a $3d$ narrow-gap insulator with a gap of approximately 30 meV[32], long studied for its colossal Seebeck effect[33] and as a candidate Kondo insulator[34–36], in which the resistivity saturates at low temperatures. Recent angle-resolved photoemission spectroscopy (ARPES)[37] and transport measurements[38] have revealed metallic surface states that emerge in this saturation regime, yet their origin has remained unsettled, and whether the surface itself is strongly correlated or magnetic has not been established.

In this structure, each Fe atom sits in a distorted octahedron of Sb atoms (Fig. 1**b**), and neighboring octahedra share edges to form chains along the $c$ axis (Fig. 1**c**). At the level of DFT, the electronic structure near the Fermi energy can be understood as follows. Taking the local axes ($x$, $y$ and $z$ axes) along the octahedral Fe–Sb bonds, the states near the Fermi energy $E_F$ mainly comprise two types of Fe $3d$ orbitals: $d_{xy}$ and $d_{xz}/d_{yz}$, which behave very differently (Fig. 1**d–f**). The $d_{xy}$ orbitals extend their lobes along the $c$ axis toward neighboring Fe atoms; their overlap produces bonding states around 1.3 eV below $E_F$ and antibonding states just above $E_F$ (Fig. 1**e**)[35,39], forming a dispersive, itinerant band along the Fe-Fe direction (*e.g.*, Γ-Z in Fig. 1**d**). The $d_{xz}$ and $d_{yz}$ orbitals extend their lobes toward neither the neighboring Fe atoms nor the Sb ligands, remaining essentially nonbonding; they form a comparatively flat band in all directions, located just below $E_F$ (Fig. 1**d**, **e**). The bonding character of these states is confirmed by the crystal orbital Hamilton population (COHP)[40–42], in which a positive (negative) value of −COHP indicates bonding (antibonding) with the magnitude measuring its strength (Fig. 1**f**; orbital-resolved COHP in Supplementary Fig. 1). The narrow gap of $FeSb_2$ thus opens between the flat, nonbonding $d_{xz}/d_{yz}$ states at the top of the valence band and the dispersive, antibonding $d_{xy}$ states at the bottom of the conduction band. Beyond this band picture, however, electron correlations are widely recognized to play an essential role in $FeSb_2$[32,43]. As one such interpretation, the hybridization between the flat and dispersive bands has been suggested to underlie this correlated behavior, recently described in terms of orbital-selective Kondo and Hund-metal physics[44,45].

To examine the surface electronic structure and its topological origin, we performed slab calculations for the (101) surface of our films, together with a Wannier charge center (WCC) analysis of the bulk (Fig. 1**g–i**). The slab calculation yields surface bands that traverse $E_F$ within the bulk gap (bold lines in Fig. 1**g**), consistent with the metallic surface states reported previously by ARPES[37] and transport measurements[38]. The corresponding wavefunctions decay into the bulk over the outermost few atomic layers and float off the surface atoms (Fig. 1**h**). To trace the origin of this floating character, we examined the WCCs of the valence electrons (Fig. 1**i**). The WCCs do not coincide with the atomic sites but lie in the interatomic region, which is consistent with ref. [23], resulting in a topological polarization of the bulk. By the bulk-edge correspondence, this displaced charge necessarily emerges at the termination as floating, polar surface states. We therefore assign these surface states of $FeSb_2$ to polar surface states of topological-polarization origin.

**Direct observation of correlation-driven orbital reconstruction**

A defining feature of strong correlations is the sensitivity of the electronic state to small perturbations. In $FeSb_2$, we find that even a moderate change in temperature drives a pronounced reconstruction of the Fe 3$d$ orbital occupation, which we resolve directly in real space. We mapped the VED at several temperatures by synchrotron X-ray diffraction, obtaining it through the core differential Fourier synthesis (CDFS) method[46,47] (Fig. 2**a**–**c**; see Methods and Supplementary Fig. 2 for the full set of temperatures). The three-dimensional (3D) maps reveal a marked change in the dense regions of the VED between 100 K and 200 K (Fig. 2**a**, **b**). The two cross-sections (cut 1 and cut 2 in Fig. 2**a**–**c**) clarify the orbital contributions to the VED. At 100 K, strong intensity appears between the $z$ axis and the $xy$-plane direction (cut 1; Fig. 2**a**), clearly revealing the $d_{xz}$ and $d_{yz}$ orbitals[48]; this matches the orbital occupation from DFT (Fig. 1**e**). As the temperature rises to 200 K, this intensity weakens (cut 1; Fig. 2**b**), while stronger intensity develops in the two quadrants perpendicular to the Fe chain (cut 2; Fig. 2**b**). This pattern indicates a growing contribution from the antibonding $d_{xy}$ orbital, which is depleted along the chain. The increasing antibonding occupation weakens the Fe-Fe bonds, consistent with the large thermal expansion along the $c$-axis[35] (Supplementary Fig. 3). By 300 K, the distribution changes little from that at 200 K, with additional weak intensity along the $z$ axis from the $d_{z^2}$ orbital (cut 1; Fig. 2**c**).

To quantify this reconstruction, we evaluated the orbital occupancies through the complex hybrid orbital decomposition (CHOD) method[49] as shown in Fig. 2**d** (see Methods). The result confirms the qualitative trend described above. Between 100 K and 200 K, the $d_{xz}/d_{yz}$ occupancy decreases, while the $d_{xy}$ occupancy increases, reversing their relative occupancy. This reconstruction, the origin of the dramatic change in VED (Fig. 2**a**–**c**), is far too large to arise from thermal excitation, since its onset temperature (~ 100 K) is well below the gap energy of 30 meV (~ 350 K). It instead suggests an origin in strong electronic correlations. Indeed, DFT+DMFT calculations show that the local spectral function (corresponding to the DOS) is redistributed and smeared over a wide energy range (approximately ±1 eV; the scale of the on-site interaction $U$) as the temperature rises (Fig. 2**e**). This redistribution substantially mixes the conduction-band antibonding $d_{xy}$ weight into the occupied states (also see Fig. 1**e**, **f**). This smeared incoherent weight, which fills the gap at high temperatures and disappears from the gap at low temperatures (insets of Fig. 2**d**), also drives the insulator-to-metal crossover observed in $FeSb_2$[34,44,45,50].

This orbital reconstruction has a direct consequence for the topological polarization, which arises from bonding charge in the interatomic region. As the antibonding $d_{xy}$ weight grows in the occupied states with increasing temperature, it is expected to deplete this bonding charge and weaken the polarization. The polarization should therefore be most pronounced at low temperatures, in conjunction with the disappearance of the incoherent weight from the gap.

**Nonreciprocal transport from the polar surface state**

The first-principles calculations predict surface states arising from the displaced WCCs of the bulk, while the bulk electronic state is strongly modulated by correlations, as seen above. The surface should therefore reflect both the topological polarization and the correlated reconstruction of the bulk. To probe the surface selectively, we fabricated EDLT devices from $FeSb_2$(101) thin films (23.6-nm thickness) grown on MgO(001) (Fig. 3**a**; see Methods and Supplementary Fig. 4 for details). The thin film limits the bulk volume to maximize the surface contribution, whereas the EDLT confines carrier doping to within approximately one nanometer of the surface[51]. A gate voltage $V_G$ above 3 V drives a pronounced drop in sheet resistance $R_{xx}$ below 100 K, reaching 56% at 5 V and 2 K (Fig. 3**b**). At 50 K, the sheet carrier density from Hall measurements rises accordingly from $n \approx 4.7 \times 10^{14}$ cm$^{-2}$ at 3 V to $n \approx 5.5 \times 10^{14}$ cm$^{-2}$ at 5 V (Fig. 3**c**). These results show that the surface conduction channel can be actively tuned by gating.

We measured the nonreciprocal transport to examine the surface state and its coupling to the bulk electronic reconstruction. A nonreciprocal response requires broken inversion symmetry and vanishes in a centrosymmetric crystal; it therefore selectively detects the emergence and evolution of the polar surface states in $FeSb_2$. Nonreciprocal conduction introduces a resistance term that depends on both the magnetic field $\boldsymbol{H}$ and current $\boldsymbol{I}$: $R = R_0[1 + \gamma(\widehat{\boldsymbol{P}} \times \boldsymbol{H}) \cdot \boldsymbol{I}]$ in a polar system[52,53]. This second term is detected as a second-harmonic sheet resistance $R_{xx}^{2f}$ under an AC current. Here, $R_0$ is the conventional sheet resistance, $\widehat{\boldsymbol{P}}$ is the unit vector of surface polarization $\boldsymbol{P}$ and $\gamma$ is the nonreciprocal coefficient. The signal is maximized when $\boldsymbol{P}$, $\boldsymbol{H}$ and $\boldsymbol{I}$ are mutually perpendicular, with $\boldsymbol{I} \parallel \boldsymbol{x}$, $\boldsymbol{H} \parallel \boldsymbol{y}$, and $\boldsymbol{P} \parallel -\boldsymbol{z}$ (Fig. 3**a**; Methods and Supplementary Fig. 5). Here, the *x*, *y*, and *z* axes are defined by the measurement geometry rather than by the Fe-Sb bonds. In this configuration, a clear $R_{xx}^{2f}$ appears below 50 K, indicating the emergence of polar surface states (Fig. 3**d**). The $R_{xx}^{2f}$ profile is linear in *H*, and the magnitude of $\gamma$, estimated from its slope, peaks around 25 K and decreases toward lower temperature (Fig. 3**e**). Over the same temperature range, the carrier density changes dramatically, reversing its sign around 100 K and then falling by two orders of magnitude (inset of Fig. 3**c**). The nonreciprocity is also sensitive to the gate voltage. Above 3 V, where the carrier density begins to rise, the peak of $\gamma$ shifts in both temperature and magnitude (Fig. 3**e**), corroborating the surface origin of the nonreciprocal signal. We note that changing the *H*-direction rules out a thermoelectric (Nernst) origin (Supplementary Fig. 6).

The gate-induced drop in resistance, the dramatic change in the carrier density, and the onset of nonreciprocity all appear below 100 K, coinciding with the temperature scale of the bulk orbital reconstruction revealed by CDFS and DFT+DMFT (Fig. 2). The steep suppression of the carrier density indicates that the bulk gap becomes well defined as the incoherent weight disappears from the gap. In this regime, the mixing of the antibonding $d_{xy}$ weight into the occupied states diminishes, and the topological polarization is expected to strengthen accordingly. The emergence of the nonreciprocal signal in the same regime supports this expectation, indicating that the polar surface states develop in step with the correlation-driven evolution of the bulk. The gate sensitivity of $\gamma$ further corroborates their surface origin. This observed bulk-to-surface connection demonstrates that the bulk-edge correspondence holds in $FeSb_2$, confirming the topological-polarization origin of the surface states assigned in Fig. 1.

**Gate-induced surface magnetic transition**

Tuning the carrier density is a canonical route to phase transitions in strongly correlated systems. At the polar surface of $FeSb_2$, gating indeed induces a magnetic transition. At gate voltages above 4 V, a pronounced negative magnetoresistance develops at low temperatures (Fig. 4**a**–**c**), indicating the suppression of spin-disorder scattering by an emerging magnetic order. Correspondingly, a clear anomalous Hall resistance $R_{yx}^{\mathrm{A}}$ emerges above 4 V (Fig. 4**d**–**f**; the total Hall resistance in the insets). At 5 V and 2 K, $R_{yx}^{\mathrm{A}}$ exhibits hysteresis (Fig. 4**d**), evidencing surface ferromagnetism with remanent magnetization. The amplitude of $R_{yx}^{\mathrm{A}}$ mapped over the temperature-gate voltage plane (Fig. 4**g**) shows the ordered region expanding with $V_G$, with the magnetic transition temperature $T_c$ reaching 50 K (see Supplementary Fig. 7 for the determination of $T_c$). At the base temperature of 2 K, the order emerges between 3 and 4 V, through a gate-induced quantum phase transition.

The magnetism cannot be attributed merely to the near-surface bulk bands. In bulk $FeSb_2$, chemical electron doping induces magnetic order above 0.2 electrons per formula unit[54–56]. The gate-induced modulation here ($\sim 5 \times 10^{13}$ $cm^{-2}$ at 4 V; Fig. 3**c**) amounts to ~ 0.02 electrons per formula unit for the active thickness of ~ 1 nm, which is several times below the bulk threshold. The magnetic order at such a low doping level instead points to the correlated surface states, which supply the high density of states required for a magnetic instability. The surface states span a bandwidth of only 200 meV with nearly dispersionless portions (*e.g.*, along $\overline{X}$-$\overline{S}$; Fig. 1**g**), much narrower than the dispersive bulk conduction band (Fig. 1**b**). Such narrow surface bands can order magnetically in two ways. In one, their large density of states satisfies the Stoner criterion as gating raises the Fermi level. In the other, their flat portions host local moments while the dispersive portions supply itinerant carriers, and gating strengthens the carrier-mediated RKKY coupling until it overcomes the Kondo screening, as in the Doniach picture of heavy-fermion systems[1,2,57]. Both routes are facilitated by the high density of states of the narrow surface bands, consistent with the low threshold observed here. Notably, the nonreciprocal response persists into the magnetic phase (Fig. 3**e**), showing that the surface states retain their polarization across the transition. The coexisting polarization and magnetic order thus strongly suggest correlation and topology acting together on a single surface.

As for the nature of the order, our first-principles calculations for the magnetically ordered surface suggest an altermagnetic spin splitting (Supplementary Note 1), in line with the proposal of altermagnetism in bulk $FeSb_2$[58]. The gate-induced order may thus be a two-dimensional altermagnetic state.

**Conclusion and outlook**

In summary, we have shown that the metallic surface states of $FeSb_2$ are polar surface states of topological-polarization origin, governed by the strong correlations of the bulk. The correlation-driven orbital reconstruction, imaged in real space by CDFS and reproduced by DFT+DMFT, modulates the topological polarization of the bulk. The nonreciprocal surface transport tracks this reconstruction, manifesting the bulk-edge correspondence, and gating finally drives the surface across a magnetic transition. These results establish a correlated topological surface, in which strong correlations control and ultimately transform a topological-polarization surface state.

Such correlated topological surfaces are not limited to $FeSb_2$. Metallic surface states have also been reported in the correlated insulator FeSi[28], and the catalogues of obstructed atomic insulators[23–25] list many candidates among 3*d* transition-metal compounds. Topological polarization requires no heavy elements, and flat-band correlations extend across *d*-electron intermetallics. Their combination reaches far beyond $FeSb_2$, charting a broad family of correlated topological materials.

These surfaces may further serve as a complementary platform for two-dimensional strongly correlated physics. Correlated oxides host correlated phases persisting to high temperatures, but their large energy scales make demand large carrier changes to switch between phases[59]; moiré superlattices are highly tunable, but their small energy scales confine the correlated phases to very low temperatures. Correlated topological surfaces lie in between, combining transition temperatures of tens of kelvin with full electrostatic tunability. Further emergent phases, such as superconductivity and density waves, await exploration at the natural boundaries of bulk crystals.

## Methods

### Bulk single-crystal growth of $FeSb_2$

High-quality $FeSb_2$ single crystals were grown from stoichiometric amounts of 99.99% pure Fe and 99.9999% pure Sb powders by the $I_2$ chemical vapor transport (CVT)[60] in a gradient of 700–650 °C in a sealed quartz tube for 11 days.

### Synchrotron X-ray diffraction measurements

Synchrotron X-ray diffraction measurements were conducted at the BL02B1 beamline of SPring-8, Japan[61]. A nitrogen-gas-blowing device was used for measurements between 100 K and 400 K, whereas a helium-gas-blowing device was used for the measurement at 50 K. Diffraction patterns were recorded using a CdTe PILATUS two-dimensional detector (dynamic range ~$10^6$) with an X-ray energy of 40 keV. The intensities of Bragg reflections corresponding to $d > 0.30$ Å were collected with the CrysAlisPro software[62] using a fine-slice method, in which the data were obtained by dividing the reciprocal space region in increments of $\omega = 0.01°$. The intensities of equivalent reflections were averaged, and the structural parameters were refined using JANA2006[63] . High-angle reflections ($\sin\theta/\lambda > 0.5$ Å) were used exclusively for structural refinement to perform high-angle analysis. Since the contribution of spatially extended valence electrons is very small in the high-angle region, the structural parameters, including the atomic displacement parameters (mainly due to the thermal vibrations), are obtained with high accuracy. See Supplementary Tables 1–5 for detailed results of the structural analysis.

### CDFS analysis

The CDFS method was used to extract the VED distribution around each atomic site[46]. The core electron configurations were assumed to be [Ar] for the Fe atom and [Kr]$4d^{10}$ for the Sb atom. The contribution of thermal vibrations was subtracted from the VED using the atomic displacement parameters determined by the high-angle analysis. The 3D VED was reconstructed on a real-space grid with a voxel size of 0.02 Å. It should be noted that the absolute value of the obtained VED does not directly reproduce the number of valence electrons around the atoms, partly because the double scattering, absorption, extinction, and detector saturation[64] could not be completely excluded in the measurements of diffraction intensities.

### CHOD analysis

The CHOD method was applied to the valence electron density distributions obtained by the CDFS method[49]. The CHOD method optimizes the electron density obtained by the CDFS method under the following formulation:

$$\text{minimize}_{P,\rho_{\text{offset}}} \frac{\int_{|\mathbf{r}-\mathbf{r}_0|<r_{\max}} w(\boldsymbol{r})\left(\rho_{\text{CDFS}}(\mathbf{r}) - \rho_{\text{rec}}(\mathbf{r})\right)^2 \mathrm{d}\mathbf{r}}{\int_{|\mathbf{r}-\mathbf{r}_0|<r_{\max}} w(\boldsymbol{r})\rho_{\text{CDFS}}^2(\mathbf{r})\mathrm{d}\mathbf{r}},$$

where

$$\rho_{\mathrm{rec}} = \frac{1}{V}\sum_{|\boldsymbol{k}|<K_{\max}} e^{i\mathbf{k}\cdot\mathbf{r}-\mathbf{k}^{\mathrm{T}}U\mathbf{k}} \int \left(\sum_{ij} P_{ij}\psi_i^*(\mathbf{r})\psi_j(\mathbf{r})\, e^{-i\mathbf{k}\cdot\mathbf{r}}\right) \mathrm{d}\mathbf{r} + \rho_{\mathrm{offset}}.$$

Here, $V$ is the volume of the unitcell, $\mathbf{r}_0$ is the position of the atomic nucleus, $r_{\max}$ ($K_{\max}$) is the cutoff radius of real (reciprocal) space, $U$ is the matrix of atomic displacement factor and we use $w(\mathbf{r}) = \rho_{\mathrm{CDFS}}^2(\mathbf{r})$ as a weight. $P_{ij}$ and $\rho_{\mathrm{offset}}$ are fitting parameters, and $P$ is a positive semidefinite matrix.

As basis functions $\psi$, 3*d* atomic orbitals were employed. Since the Fe site possesses spatial inversion symmetry, the *p*-orbitals do not hybridize with the *d*-orbitals. In this case, the diagonal elements of $P_{ij}$ are proportional to the population of each orbital. To account for differences in local environments, a magnification factor $\kappa$ was introduced, defining $\psi(\mathbf{r}) = \frac{1}{\kappa^3}\, \psi_{\mathrm{original}}\left(\frac{\mathbf{r}}{\kappa}\right)$. The axis orientation was defined with Fig. 1**b**, and optimization was performed within a radius of 0.6 Å from the nucleus. The error of the solution was derived using the variance-covariance matrix.

**Thin-film growth of $FeSb_2$**

The $FeSb_2$(101) thin films were grown on a MgO(001) substrate by molecular beam epitaxy (MBE). The substrates were degassed at 500 °C for several hours prior to growth. Fe and Sb were co-evaporated onto the substrate at 500 °C, with the flux ratio of Fe:Sb set to approximately 1:12 to compensate for the re-evaporation of Sb from the substrate surface due to its high vapor pressure. The films were subsequently annealed at 610 °C. The film quality and thickness were confirmed by *θ-2θ* X-ray diffraction (Supplementary Fig. 3).

**Device fabrication**

The ion-gating devices used in this study were fabricated by a standard process without lithography as follows. 10-nm-thick Ti and 50-nm-thick Au were deposited onto the films as source/drain electrodes and voltage probes by an electron-beam evaporator through a metal mask with a Hall-bar pattern. The films were then mechanically scratched between the electrodes to isolate the voltage probes and to define a channel size to be 100 μm in length and 300 μm in width. Subsequently, a Pt plate was placed above the channel region as a gate electrode, and a droplet of the ionic liquid was inserted in between the channel and the Pt plate just before the measurements. The ionic liquids used in this study was N,Ndiethyl-N-(2-methoxyethyl)-N-methylammonium bis-trifluoromethylsulfonyl-imid ([DEME][TFSI]).

**Transport measurements**

The AC transport properties of the ion-gating devices were characterized using a combination of a current source (Keithley 6221) and lock-in amplifiers (NF LI5650 and Stanford Research Systems SR830). The DC gate voltage $V_G$ was applied using a source measure unit (Keithley 2450) at 220 K, after which the sample was cooled to the target temperature for measurements. All measurements were performed in a Physical Property Measurement System (PPMS, Quantum Design).

The nonreciprocal conduction was evaluated by measuring the second-harmonic signal of the AC sheet resistance $R_{xx}^{2f}$ under an in-plane magnetic field $\boldsymbol{H}$ applied perpendicular to the current direction. When an AC input current $I = I_0 \sin(2\pi f t)$ was applied, the second-harmonic sheet resistance is expressed as $R_{xx}^{2f} = -\frac{1}{2} R_0 \gamma I_0 \mu_0 H$. The current frequency was 13 Hz and the current amplitude was typically 93 μA, which corresponds to a current density of $2.0 \times 10^7$ A/m$^2$ in the film. The linear current dependence of $R_{xx}^{2f}$ was confirmed (Supplementary Fig. 12).

Reproducibility was checked on a separately grown film (film B; Supplementary Figs. 13, 14). The gate-induced drop in resistance, the increase in the carrier density, the emergence of the nonreciprocal transport below 50 K, and the emergence of the anomalous Hall effect are all reproduced. In film B, however, the nonreciprocal signal reverses its sign upon gating; the origin of this sign reversal remains to be clarified.

**First-principles calculations**

Bulk band structure calculations were performed using the Vienna Ab initio Simulation Package (VASP) code[65] with the projector augmented wave (PAW) method[66]. The Perdew-Burke-Ernzerhof (PBE) generalized gradient approximation (GGA) was employed for the exchange-correlation functional[67]. A 9×8×16 k-point mesh and a 400 eV planewave energy cutoff were used, and spin-orbit coupling (SOC) was included. The crystal structure was taken from the experimental high-angle XRD data at 100 K. The projected density of states (pDOS) and the COHP were calculated with the LOBSTER package[42] without SOC.

Slab band structure calculations were performed using VASP with the same PBE functional and energy cutoff. The crystal structure was taken from ICSD-42604[68] in the Inorganic Crystal Structure Database (ICSD)[69]. A slab model of 60 atoms (20 Fe atoms and 40 Sb atoms) was constructed in an orthogonal unit cell, with the *b* axis aligned along the bulk *b*-axis and the *c* axis along the (101) surface normal. A vacuum layer of 34 Å was added along the *c* axis, and the Brillouin zone was sampled with a 6×6×1 k-mesh. The atomic positions were relaxed with the cell shape and volume fixed without SOC, after which SOC was included in the self-consistent field (SCF) calculations.

The Wannier-function analysis was performed using Quantum ESPRESSO[70] and Wannier90[71] in combination with symWannier[72]. The PBE functional and PAW pseudopotentials were used without SOC. The wave-function and charge-density cutoffs were set to 46 Ry and 237 Ry, respectively. Both SCF and NSCF calculations were performed on a 6×6×12 k-mesh. Wannier functions for the Fe 3*d* and Sb 5*p* orbitals were constructed using a frozen energy window extending to 2.0 eV above the Fermi level. The WCC calculations were performed following the numerical procedure implemented in WannierTools[73].

For the DFT+DMFT calculations, the DFT part was treated with WIEN2k code[74] within the PBE functional, using 12×10×22 k-mesh. The DMFT impurity problem is solved by the hybridization-expansion continuous-time quantum Monte Carlo (CTQMC) method, in a projection-based DFT+DMFT setup with full charge self-consistency[31,75,76]. The on-site Coulomb interaction and the Hund coupling were set to $U$ = 5.0 eV and J = 0.8 eV, respectively.

**Acknowledgements** We thank T. Arima, R. Arita and C. Koyama for experimental and theoretical supports. This work was supported by JSPS KAKENHI (Grants No. 23H04017, 23H04869, 23H05431, 23H05462, 24H00417, 24H01212, 24H01644, 24H01652, 25H02126, 26H00626, 26H00633, 26K00652, 26KJ0993), JST FOREST (Grants No. JPMJFR2038, No. JPMJFR2362), JST CREST (Grant No. JPMJCR23O3), JST ASPIRE (Grants No. JPMJAP2317, No. JPMJAP2512), PRESTO (Grants No. JPMJPR25H9), the Samco Foundation. The synchrotron radiation experiments were performed at SPring-8 with the approval of the Japan Synchrotron Radiation Research Institute (JASRI) (Proposal No. 2024B2017, 2025A1505, 2025A1998, 2025B1928, 2025B2178, 2026A1683 and 2026A1684).

**Author contributions** T.I. and N.K. conceived the project. T.I. and N.K. grew the bulk single crystals and thin films of $FeSb_2$. T.I. and H.M. performed transport experiments. T.I., S.A., S.K., and Y.N. performed XRD measurements. T.I. performed the first-principles calculations on bulk and slab setups under the supervision of T.K. G.H. and K.K. performed the Wannier function analysis and the DFT+DMFT calculations under the supervision of T.K. N.K. organized the project. T.I. and N.K. wrote the draft and all the authors discussed the results and commented on the manuscript.

**Competing interests** The authors declare no competing interests.

**Data availability** All the data presented in figures are available at UTokyo Repository: https://...

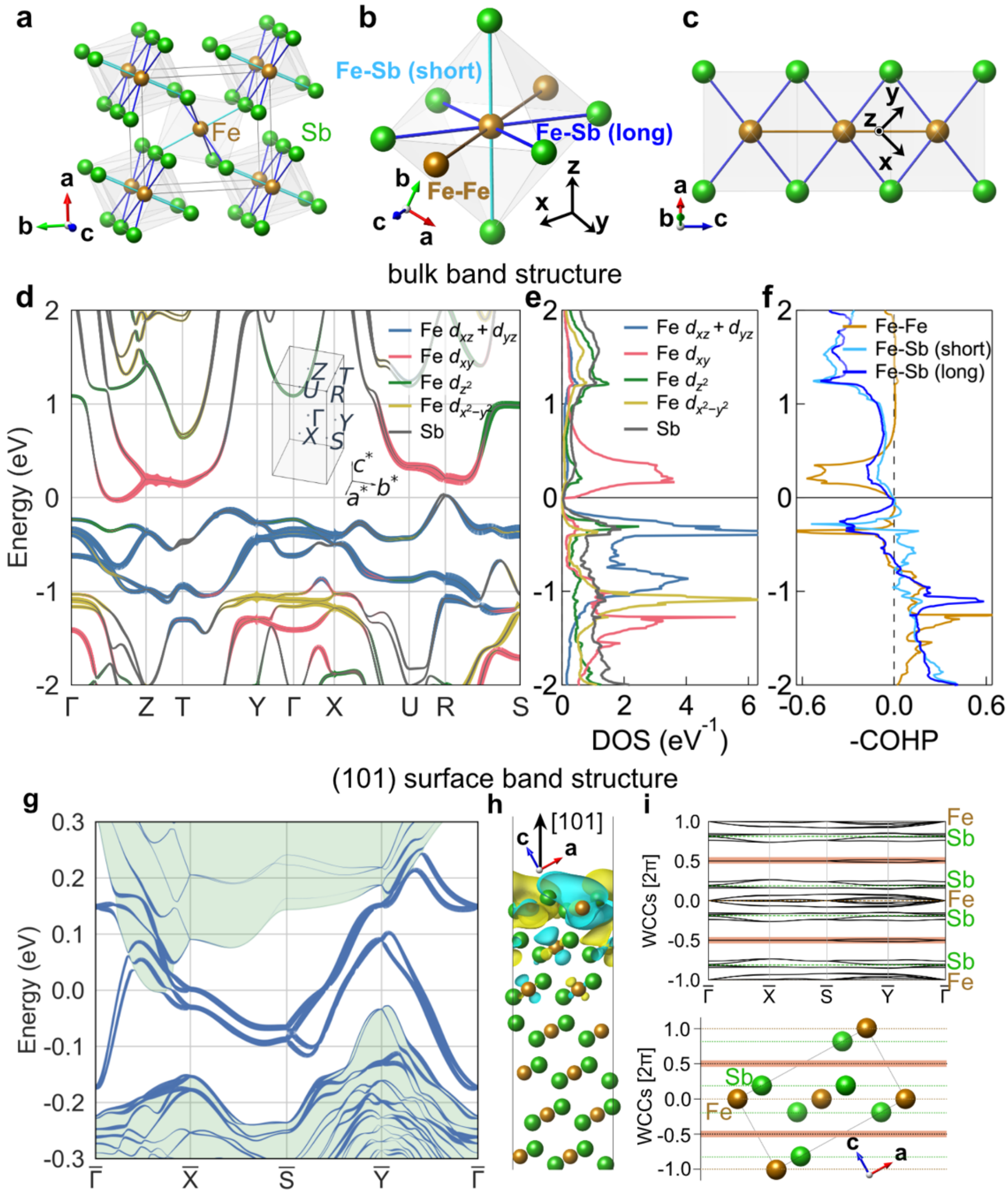

**Fig. 1 | Bulk and surface electronic states of $FeSb_2$ from first-principles calculations.**

**a–c,** Crystal structure of $FeSb_2$, drawn using VESTA: the unit cell (**a**); the local octahedral coordination of Fe (**b**); and the edge-sharing octahedral chain along the *c* axis with the local axes (*x*, *y*, *z*) (**c**). Short Fe-Sb, long Fe-Sb, and Fe-Fe bonds are colored light blue, blue, and brown, respectively. **d–f,** Bulk electronic states: the band structure (**d**); the projected density of states (pDOS) (**e**); and the crystal orbital Hamilton population (COHP) for the three bonds defined in **a–c** (**f**). In **d** and **e**, blue, red, green, yellow, and gray denote the Fe $d_{xz}/d_{yz}$, $d_{xy}$, $d_{z^2}$, $d_{x^2-y^2}$, and Sb *p* orbital contributions, respectively. **g, h,** Surface electronic states from the slab calculation for the (101) surface: the band structure (**g**), in which the line thickness represents the weight of the outermost 3 atomic layers and the green shading marks the bulk bands; and the real-space distribution of the surface-state wavefunction at the Γ point (**h**). **i,** Wannier charge center (WCC) evolution along the surface Brillouin zone, calculated using the Wilson loop method (top), and the corresponding WCC position along the [101] direction, shown within the crystal structure (bottom). Brown and green lines indicate the positions of Fe and Sb atoms, respectively. The red lines mark WCCs located at interatomic positions, which give rise to the surface states.

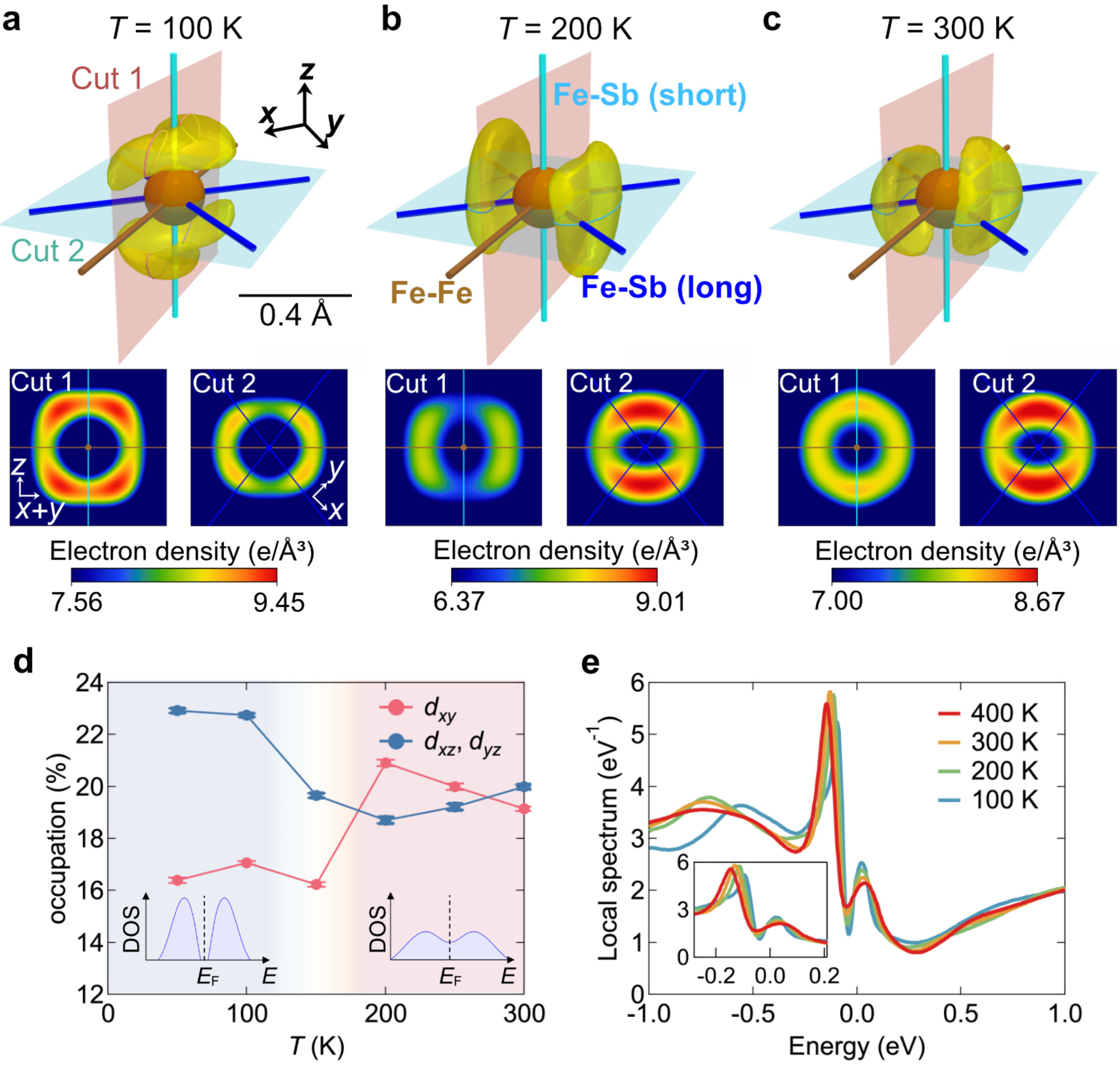

**Fig. 2 | Direct observation of the correlation-driven orbital reconstruction in $FeSb_2$.**

**a–c,** Valence electron density (VED) around the Fe site, obtained by the CDFS method, at 100 K (**a**), 200 K (**b**), and 300 K (**c**). The upper panels show VED isosurfaces at 9.15, 8.3, and 8.3 $e$/Å$^3$, respectively. Light blue, blue, and brown lines indicate the directions of the short Fe-Sb, long Fe-Sb, and Fe-Fe bonds, respectively. The two cross-sections (cut 1 and cut 2) are indicated in **a**. The lower panels show VED color maps on cut 1 (left) and cut 2 (right), with the bond directions indicated by lines. **d,** Temperature dependence of the orbital occupations from the CHOD analysis. The insets show schematics of the DOS in the two regimes: a well-defined gap with coherent quasiparticle peaks at low temperatures (left) and a smeared, incoherent DOS filling the gap at elevated temperatures (right), corresponding to the spectra in the inset of **e**. **e,** DFT+DMFT local spectra of $FeSb_2$ at selected temperatures. The spectral weight is redistributed over a wide energy range of about ±1 eV with increasing temperature, driving the reconstruction of the orbital occupation. The inset magnifies the spectra around $E_F$, showing that the gap is progressively filled with increasing temperature.

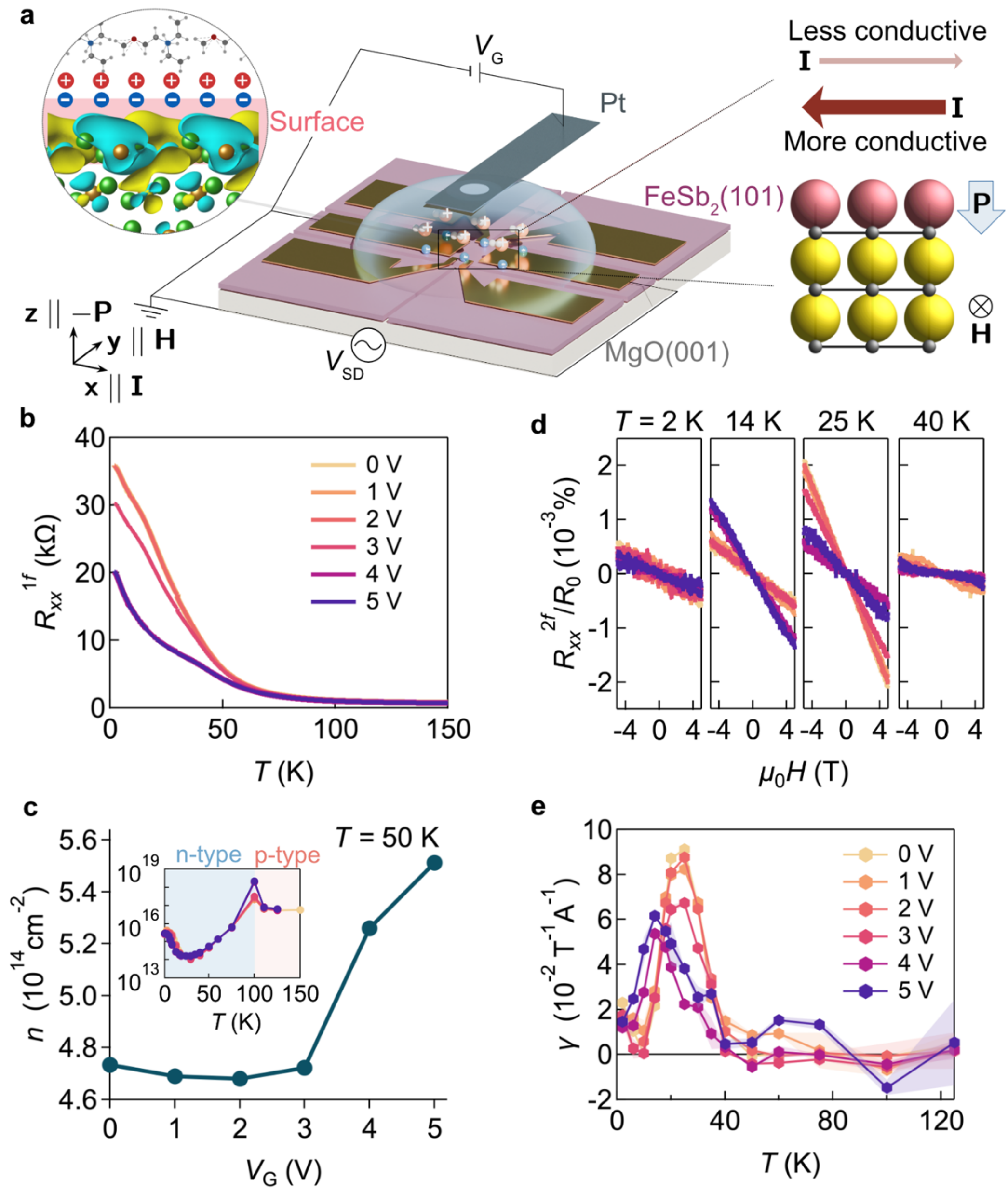

**Fig. 3 | Detection of the polar surface states by nonreciprocal transport in an EDLT device.**

**a,** Schematic of the EDLT device. The left inset shows an enlarged view of the interface region, illustrating carrier doping at the surface of the film hosting the topological-polarization-derived surface states. The right inset illustrates the origin of the nonreciprocal transport: the interatomic charge of the topological polarization (yellow) protrudes above the topmost atoms at the surface, forming a surface polarization ***P*** (red) pointing downward. Arrows in the main panel indicate the configuration used for transport measurements: current ***I*** || ***x***, magnetic field ***H*** || ***y***, and ***P*** || -***z***. Here, the *x*, *y*, and *z* axes are defined by the measurement geometry rather than by the Fe-Sb bonds. In this polar configuration, the conduction depends on the current direction (arrow thickness), giving rise to the nonreciprocal response. **b,** First-harmonic sheet resistance $R_{xx}^{1f}$ as a function of temperature at gate voltages from 0 V to 5 V. **c,** Sheet carrier density, extracted from Hall measurements, as a function of gate voltage at 50 K. The inset shows its temperature dependence at the same gate voltages, colored as in **b**. **d,** Second-harmonic sheet resistance $R_{xx}^{2f}$, normalized by the zero-field sheet resistance $R_0$, as a function of magnetic field $\mu_0 H$ at 2 K, 14 K, 25 K, and 50 K, and at gate voltages from 0 V to 5 V, colored as in **b**. **e,** Nonreciprocal coefficient $\gamma$ as a function of temperature at gate voltages from 0 V to 5 V. The shaded areas represent the measurement errors.

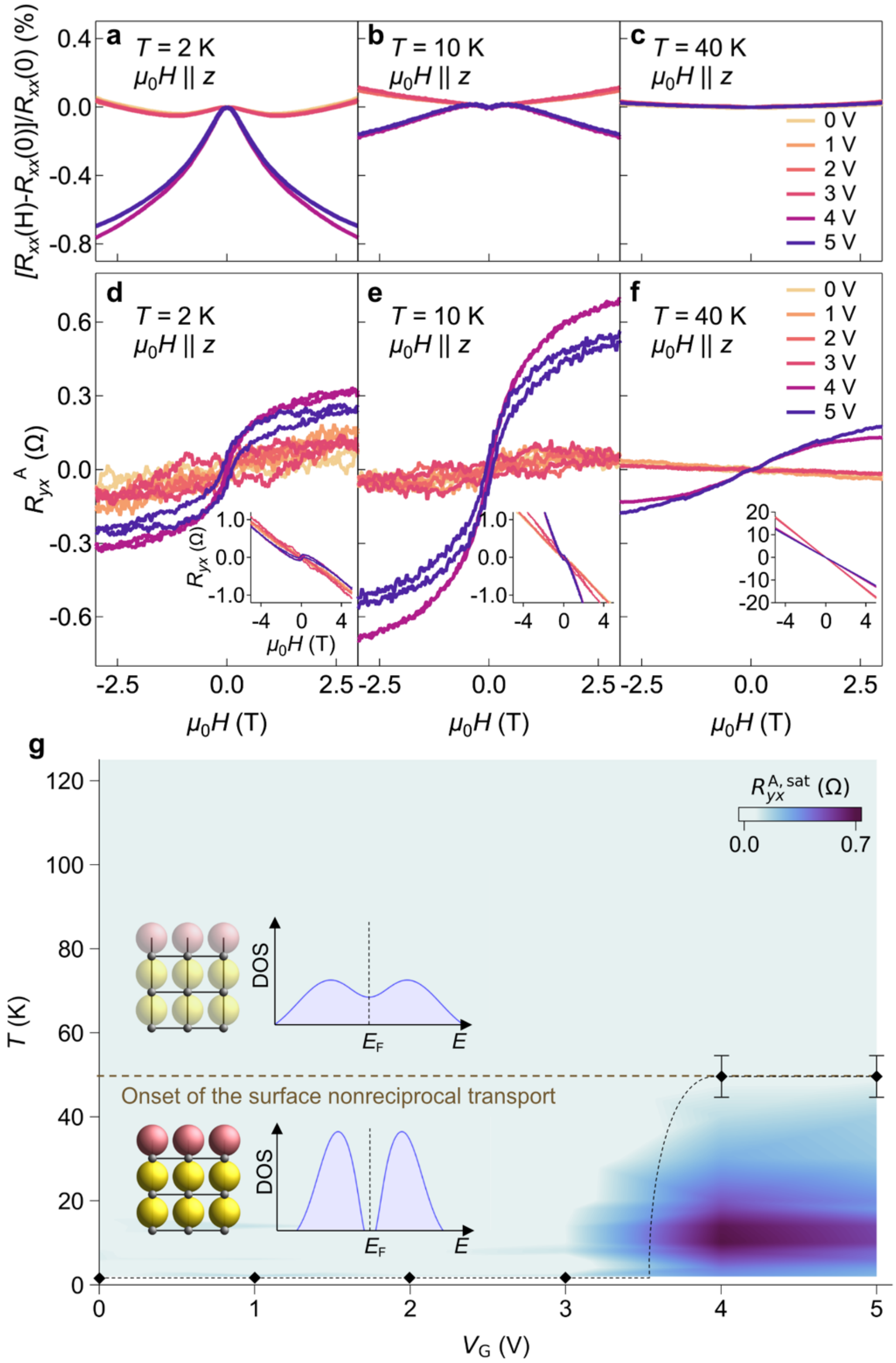

**Fig. 4 | Gate-induced surface magnetic transition.**

**a-c,** Magnetoresistance $[R_{xx}^{1f}(H) - R_{xx}^{1f}(0)]/R_{xx}^{1f}(0)$ as a function of magnetic field at 2 K (**a**), 10 K (**b**), and 30 K (**c**), and at gate voltages from 0 V to 5 V. **d–f,** Anomalous Hall resistance $R_{yx}^{\mathrm{A}}$, obtained by subtracting a linear fit to the high-field regions (3 T < $|\mu_0 H|$ < 5 T) from the raw Hall resistance, as a function of magnetic field at the same temperatures and gate voltages as in **a–c**. The insets show the raw Hall resistance $R_{yx}$. **g,** Gate voltage–temperature phase diagram with the saturated anomalous Hall resistance $R_{yx}^{\mathrm{A,sat}}$ shown as a color map. The points with error bars indicate the transition temperature $T_c$ at each gate voltage (see Supplementary Fig. 7 for the determination of $T_c$). The horizontal dashed line marks the onset temperature of the surface nonreciprocal conduction. The schematics within the phase diagram illustrate the evolution of the topological polarization (yellow) and the surface polarization (red): at high temperatures, both are diminished by the antibonding admixture (translucent colors); at low temperatures, both are fully developed (solid colors).